# The evolving networks of debtor-creditor relationships with addition and deletion of nodes: a case of P2P lending

Lin Chen, Ping Li, Qiang Li

(School of Management and Economics, University of Electronic Science and Technology of China, Chengdu, China, 611731)

**Abstract**: P2P lending activities have grown rapidly and have caused the huge and complex networks of debtor-creditor relationships. The aim of this study was to study the underlying structural characteristics of networks formed by debtor-creditor relationships. According attributes of P2P lending, this paper model the networks of debtor-creditor relationships as an evolving networks with addition and deletion of nodes. It was found that networks of debtor-creditor relationships are scale-free networks. Moreover, the exponent of power-law was calculated by an empirical study. In addition, this paper study what factors impact on the exponent of power-law besides the number of nodes. It was found that the both interest rate and term have significantly influence on the exponent of power-law. Interest rate is negatively correlated with the exponent of power-law and term is positively correlated with the exponent of power-law. Our results enriches the application of complex networks.

*Key words*：P2P lending; networks of debtor-creditor relationships; degree distribution; exponent of Power-law; preferential attachment

## 1 Introduction

P2P(Peer-to-Peer) lending is a type of social lending where lenders and borrowers can do business without the help of institutional intermediaries such as banks. The concept is not new, but it is basically an individual, who is not a financial institution, lending money to another individual. A potential borrower lists a loan application with term, interest rate and principal on website of P2P lending platform. Individual lenders invest their money among the wide selection of loans listing on website. Then a new type of networks of debtor-creditor relationships is formed by this behavioral mechanism. The networks of debtor-creditor relationships are based on individuals without financial institutions. On the other hand, this behavioral mechanism makes P2P lending different from traditional social lending which is limited to friends or relatives, because most lenders and borrowers of P2P lending do not know each other, and P2P lending activities beyond the limits of geographical space, cultural and religious.

In recent years, P2P lending activities have grown rapidly. For example, As of September 30, 2017, the largest P2P lending platform Lending club has facilitated $31.2 billion in loan originations since it beginning operations in 2007. The number of unique borrowers on Lending club increased over the prior year by 104%, 125%, 151%, 116% and 96% for the years ending December 31, 2011, 2012, 2013, 2014 and 2015, respectively. The number of unique lenders on Lending club increased over the prior year by 32%, 37%, 44%, 28% and 39 %, for the years ending December 31, 2011, 2012, 2013, 2014 and 2015, respectively. This is significant for facilitating more efficient deployment of capital and improve the economy. Because the ability of individuals and small businesses to access affordable credit is essential to stimulating and sustaining a healthy, diverse and innovative economy, when traditional banks have higher fixed costs of underwriting and servicing, are ill-suited to meet individuals s and small business demand for small balance loans. Before the emergence of P2P lending, most of individual lenders generally lack the size and access to invest in structured products directly

and are unable to invest in individual and small business credit in a meaningful way, but P2P lending activities have changed this situation.

A large and growing number of recent papers studied contagion via interbank markets and OTC markets by network theory. Allen and Gale (2000) and Freixas, Parigi, and Rochet (2000) are among the first to study contagion in financial networks [1-2]. Other for example Leitner (2005), Mistrulli (2011), Lenzu and Tedeschi (2012), Paltalidis, Gounopoulos and Kizys(2015), Glasserman and Young(2015), Georg(2013), Lux(2016) and so on[3-9]. While interbank networks have been extensively analyzed, the networks of individual debtor-creditor relationships are much less studied. The reason may be that the scale of networks of individual debtor-creditor relationship limited to friends or relative was small in past, and the impact of default risk of individual debt is not alike the default risk of commercial banks in the financial market.

The emergence of the P2P lending provides the basis of F and make it is possible for researching large-scale networks of individual debtor-creditor relationships. Because there are more than thousands of thousands individuals involved in the loan originations through website of P2P lending platform. For example, "Yirendai" listed on NYSE, one of Chinese P2P lending platform, in the full year of 2017, it facilitated $6,364.0 million of loans to 649,154 qualified individual borrowers through its online marketplace, and facilitated 592,642 lenders with total investment amount of $7,388.9 million[1]. In other words, this also means that "Yirendai" has created a huge networks of debtor-creditor relationships between about 650,000 borrowers and 600,000 lenders in 2017. As a result, the scale of networks of debtor-creditor relationships formed by P2P lending is huge, and it is more complicated. Understanding the characteristics of individual debtor-creditor relationship networks is helpful for lending platform to control risks and to attract more borrowers having high credit score. Because P2P lending platform generates revenue from transaction and servicing fees of borrowers and lenders. P2P lending exhibits network effects that are driven by the number of participants and investments enabled through the lending platform. If many borrowers default and lending platform are unable to maintain borrower and lender trust, there will be a decline in the network effects of lending platform or in satisfactions of borrowers and lenders. At the same time, when borrowers and lenders have to search for a new lending platform for matching finance, the cost will be increasing. In short, the stable and growth debtor-creditor relationship networks are not only important for a lending platform but also for both borrowers and lenders. A lending platform make great efforts to attract marketplace participants continuously, when there are new loan originations means the debtor-creditor relationship networks are increasing. On the other hand, if the borrowers have paid off the principal and interest on maturity, the debtor-creditor relationships between borrowers and lenders are cleaned, which means deletion of edges or nodes from the networks of debtor-creditor relationships. In short, the networks of debtor-creditor relationships could be regarded as dynamic social networks.

In the past decades, it has witnessed a great development of studying and understanding the modern theory of complex networks. Some earlier studies mainly focused on foundation of the random network theory and have shown that different networks generally share some common topological properties such as small-world effects (Watts and Strogatz, 1998) [10], preferential attachment and scale-free features (Barabási and Albert, 1999) [11]. Other studies have found fruitful applications in fields as diverse as the mobile phone networks (Blondel, Guillaume, Lambiotte and Lefebvre,2006)[12], Internet (Yook, Jeong and Barabási,2002)[13], biological interacting(Ravasz, Somera and Mongru etc,2002) [14], citation networks(Goldberg, Anthony and Evans,2015)[15], scientific collaboration

---
[1] Yirendai Reports Fourth Quarter and Full Year 2017 Financial Results

networks(Ding,2011, Wallace, Larivie`re and Gingras,2012)[16-17], movie actor networks(Herr, Ke, Hardy and Börner,2007) [18], project collaboration networks(Liu, Han and Xu,2015) [19] and many social networks. The study of social networks has been traditionally hindered by the small size of the networks considered and the difficulties in the process of data collection, and data was collected usually from questionnaires or interviews (Ramasco, Dorogovtsev and Pastor-Satorras 2004) [20]. In the past few years, the computerization of data acquisition and the availability of high computing power make it is possible to study complex topology of networks of debtor-creditor relationships, which have largely been neglected in the studies of traditional disciplines of social networks.

In this paper, we model the networks of debtor-creditor relationships as an evolving networks with addition and deletion of nodes. Our research is related to the work of Moore, Ghoshal and Newman (2006) [21], their studies provided a theoretical framework. This paper contributes to the literature in three respects. First, we provide an important case of social networks with addition and deletion of nodes in real-world networks by P2P lending activities. Second, we provide an empirical study of networks of debtor-creditor relationships based on the complex network theory, and we find that networks of debtor-creditor relationships are scale-free networks, finally we get the range of the exponent of power-law. Third, we study what factors impact on the exponent of power-law besides the number of nodes, and we find that the both interest rate and term significantly influence the exponent of power-law.

The rest of this paper is organized as follows. Section 2 describes how to form the networks of debtor-creditor relationships by P2P lending activities, and model the networks of debtor-creditor relationships based on addition and deletion of nodes. Section 3 describes the data, defines the sample, and provides a simple summary of our studies. Section 4 is a short conclusion.

## 2 Networks of debtor-creditor relationships

### 2.1 Debtor-creditor relationships of P2P lending

Firstly, Let us to define what is the networks of debtor-creditor relationships of P2P lending by a simple example case in Fig.1. Suppose that time was at $t=0$, one of borrowers listed his loan application on website of P2P lending platform such as loan amount was $3000, term was 3 months and interest rate was 6%. We denoted this borrower by 'Borrower$_1$'. After that, four lenders ('Lender$_i$', $i=1, 2, 3, 4$) have lent 'Borrower$_1$' $1000, $500, $600 and $900 respectively. Then the networks of debtor-creditor relationships between 'Borrower$_1$' and 'Lender$_i$' ($i=1, 2, 3, 4$) were formed. In addition, suppose that time was at $t=1$, a new borrower ('Borrower$_2$') listed his loan application on website of P2P lending platform. A new lender ('Lender$_5$') and senior lenders ('Lender$_3$' and 'Lender$_4$') have lent 'Borrower$_2$' respectively. Above process showed the growth of networks of debtor-creditor relationships for P2P lending activities.

On the other hand, as it was shown in Fig.2. Suppose that time was at $t=T$ after one month, a loan of 'Borrower$_2$' was expired, and 'Borrower$_2$' had only this loan. When 'Borrower$_2$' have paid off principal and interest rate on time, the debtor-creditor relationships between 'Borrower$_2$' and other lenders who lent 'Borrower$_2$' money were cleaned. 'Borrower$_2$' also was removed from the networks of debtor-creditor relationship. At the same time, there was a new borrower ('Borrower$_k$') that listed his loan application on website of P2P lending platform, and two new lenders (Lender$_i$ and Lender$_j$) and other senior lenders have lent 'Borrower$_k$' money respectively. In short, Fig.2 not only showed the deletion of debtor-creditor relationships because of an expired loan, but also shows the addition of new debtor-creditor relationships.

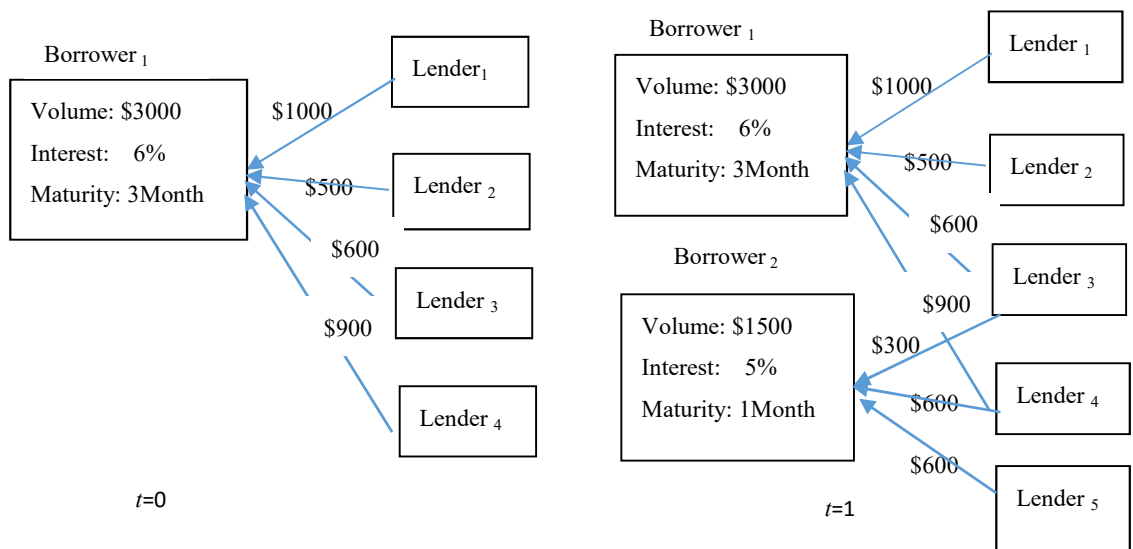

Fig.1 An example of debtor-creditor relationships

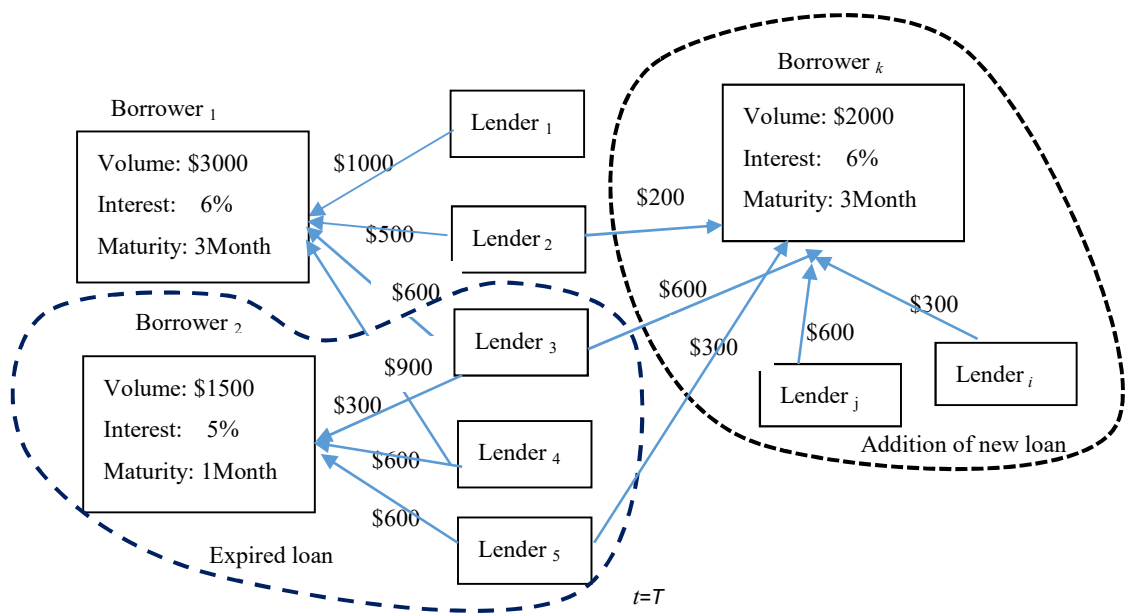

Fig.2 An example of P2P lending with an expired loan and addition of a new loan

According the formation of the networks of debtor-creditor relationships in Fig.1 and Fig.2. We denote the borrowers and lenders by the nodes and denote the debtor-creditor relationships by the edges in a graph. Because this paper only focus on the debtor-creditor relationships between borrowers and lenders, it don't need to discriminate who is debtor or creditor. As a result, the debtor-creditor relationships in Fig. 1 and Fig.2 can be denoted by the undirected networks in Fig.3.

Based on above definitions, the aim of this paper is to explore the evolution of networks of debtor-creditor relationships with deletions of nodes because of expired loans, and with additions of nodes because of new loan origination.

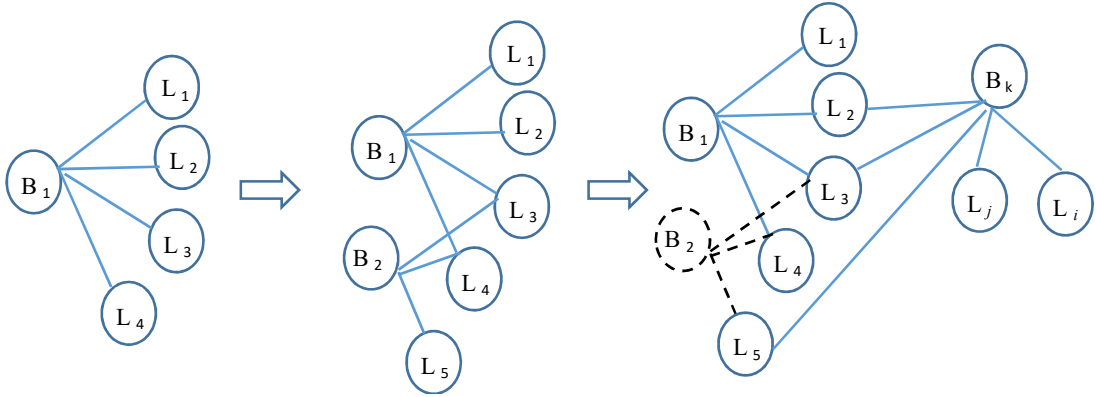

Fig.3 The evolution of network of debtor-creditor relationships

**2.2 Evolving networks of debtor-Creditor relationships**

Suppose that at time $t_0$, the networks of debtor-creditor relationships consists of $n$ vertices and $m$ edges. A node represents a lender or a borrower, and an edge represents the debtor-creditor relationship between a lender and a borrower. The degree $k_i$ of node $i$ means that node $i$ had $k_i$ debtor-creditor relationships with other nodes. Let $p_k$ denotes the fraction of nodes that have degree $k$ at a given time. By definition, there is the normalization:

$$\sum_{k=1}^{\infty} p_k = 1 \quad (1)$$

And the average degree $\langle k \rangle$ of the networks of debtor-creditor relationships is:

$$\langle k \rangle = \sum_{k=1}^{\infty} k p_k \quad (2)$$

$\langle k \rangle$ represents the mean number of debtor-creditor relationships between lenders and borrowers. For the networks of debtor-creditor relationships with $n$ nodes and $m$ edges, $\langle k \rangle = \frac{2m}{n}$.

An edge that represents a debtor-creditor relationship will be deleted from the networks of debtor-creditor relationships when the loan expires and borrower pays off principal and interest. If a node that represents a lender or a borrower has no debtor-creditor relationships with other nodes, then the node will be deleted from the networks of debtor-creditor relationships. At any time, the number of loans that will be expired is known according to the term of each loan, so the number of nodes that will be deleted from the network is known also. But the number of new debtor-creditor relationships that will be added to the network is not known, and ye it is not known that how many new nodes that represents a lender or a borrower will be added to the network of debtor-creditor relationships. Then we make the assumptions as following. In each unit of time, one node is deleted from the networks of debtor-creditor relationships because of some expired loans, and $r$ nodes are added to the networks of debtor-creditor relationships because of some new loan originations. In addition, when a new node is added to the networks of debtor-creditor relationships, there are $c$ edges are added to the network. For example, if a new node is a borrower, there are $c$ lenders to lend him (her) money. If a new node is a lender, he (she) will lend c borrowers for the aim of risk dispersion. Above assumptions was based the properties of P2P lending, and our assumptions are different from the assumptions of Moore et al.

(2006) that a single node is added and r nodes are deleted in each unit of time.

In each unit of time, there are $r$-1 nodes that are added to the networks of debtor-creditor relationships and $rc - \langle k \rangle$ edges that are added to the networks of debtor-creditor relationships. After $t$ unit of time, the number of nodes that are added to the networks of debtor-creditor relationships are $n^* = (r-1)t$, the number of new edges that are added to the networks of debtor-creditor relationships are $m^* = (rc - \langle k \rangle)t$, the average degree of new added nodes is $\langle k \rangle^*$:

$$\langle k \rangle^* = \frac{2m^*}{n^*} = 2\frac{(rc-\langle k \rangle)t}{(r-1)t} = 2\frac{rc-\langle k \rangle}{r-1} \qquad (3)$$

Let $\langle k \rangle^* = \langle k \rangle$, then get

$$\langle k \rangle = \frac{2rc}{1+r} \qquad (4)$$

If $r = 1$, the number of added nodes is equal to the number of deleted nodes, the scale of the networks of debtor-creditor relationships is constant, and there is $\langle k \rangle = c$.

In order to study that how a new borrower or lender chooses $c$ other lenders or borrowers for investing or borrowing. Let $\pi_k$ denotes the total probability that the given edge attaches to any node of degree $k$, since each edge must attach to a node, this also immediately implies that the correct normalization for $\pi_k$ is (Moore ,2006):

$$\sum_{k=1}^{\infty} \pi_k p_k = 1 \qquad (5)$$

According to the method described by Moore et al. (2006), given at time $t$, there were $n$ nodes in the debtor-creditor relationships network. The number of nodes with degree $k$ is $np_k$. The next unit of time this number was $(n + r - 1)p'_k$. The variations of this number was influenced by addition and deletion of nodes. If a node with degree $k$-1 connected to an added node with $c$ edges, the number of nodes with degree $k$ will be increased by $c\pi_{k-1}p_{k-1}$. If a node with degree $k$ connected to an added node, the number of nodes with degree $k$ will be decreased by $c\pi_k p_k$. In short, when $r$ new nodes were added in the network of debtor-creditor relationships, the variations of number of nodes with degree $k$ is:

$$r\delta_{kc} + rc\pi_{k-1}p_{k-1} - rc\pi_k p_k \qquad (6)$$

In Eq.(6), $\delta_{kc}$ was defined by:

$$\delta_{kc} = \begin{cases} 1, k = c \\ 0, k \neq c \end{cases} \qquad (7)$$

On the other hand, if a node was removed from network of debtor-creditor relationships, the degree of neighbors of removed node with degree $k$+1 was changed to $k$, the number of nodes with degree $k$ will be increased by $(k+1)p_{k+1}$, the degree of neighbors of removed node with degree $k$ was changed to $k$-1, the number of nodes with degree $k$ will be decreased by $kp_k$. If the degree of removed node was $k$, the number of nodes with degree $k$ will be decreased by $p_k$. In short, the variations of number of nodes with degree $k$ is

$$(k+1)p_{k+1} - kp_k - p_k \qquad (8)$$

Finally, According to the Eqs.(6) and (8), the evolution of the number of nodes with degree $k$ is governed by a rate equation as follows:

$$(n + r - 1)p'_k = np_k + r\delta_{kc} + rc\pi_{k-1}p_{k-1} - rc\pi_k p_k + (k+1)p_{k+1} - kp_k - p_k \qquad (9)$$

Setting $p'_k = p_k$, the Eq.(9) is equal to:

$$0 = r\delta_{kc} + rc\pi_{k-1}p_{k-1} - p_k(r + rc\pi_k + k) + (k+1)p_{k+1} \qquad (10)$$

Then both sides of Eq.(10) multiplied by $z^k$:

$$0 = r\delta_{kc}z^k + rc\pi_{k-1}p_{k-1}z^k - p_k(r + rc\pi_k + k)z^k + (k+1)p_{k+1}z^k \quad (11)$$

Summing Eq.(11) over $k(k = 0,1,2,\cdots,\infty)$:

$$0 = r\delta_{kc}\sum_{k=0}^{\infty}z^k + rc\sum_{k=0}^{\infty}\pi_{k-1}p_{k-1}z^k - \sum_{k=0}^{\infty}p_k(r + rc\pi_k + k)z^k + \sum_{k=0}^{\infty}p_{k+1}(k+1)z^k \quad (12)$$

Let us denote the generating functions of $p_k$ and $\pi_k p_k$ as following:

$$f(z) = \sum_{k=0}^{\infty}\pi_k p_k z^k \quad (13)$$

$$g(z) = \sum_{k=0}^{\infty}p_k z^k \quad (14)$$

Substituting Eqs.(13) and (14) into Eq. (12), then get:

$$0 = rZ^c + rcf(z)(z-1) - rg(z) - (z-1)\frac{\partial g}{\partial z} \quad (15)$$

The "preferential attachment" mechanism means that a node connect to others in proportion to their degree (Barabási and Albert, 1999). This mechanism is same as the Matthew effect of P2P lending market. For example, a new added lender may be more willing to lend those borrowers who had borrowed money successfully from others because of their good credit rating, and a new added borrower may be got a loan from those lenders who had lent many borrowers. Then suppose that the probability of an added edge attaching to a node with degree $k$ was proportional to $k$, then setting $\pi_k = Ak$ with constant $A$.

$$\sum_{k=1}^{\infty}\pi_k p_k = \sum_{k=1}^{\infty}Akp_k = A\sum_{k=1}^{\infty}kp_k = 1$$

$$A = \frac{1}{\sum_{k=1}^{\infty}kp_k} = \frac{1}{\langle k \rangle}$$

$$\pi_k = Ak = \frac{k}{\langle k \rangle}$$

$$f(z) = \sum_{k=0}^{\infty}\frac{k}{\langle k \rangle}p_k z^k = \frac{1}{\langle k \rangle}\sum_{k=0}^{\infty}kp_k z^k = \frac{z}{\langle k \rangle}\sum_{k=0}^{\infty}kp_k z^{k-1} = \frac{z}{\langle k \rangle}\frac{\partial g}{\partial z} \quad (16)$$

Substituting Eq.(16) into Eq. (15), then get:

$$\frac{\partial g}{\partial z}\left[(1-z)\left(rc\frac{z}{\langle k \rangle} - 1\right)\right] + rg(z) -= rz^c \quad (17)$$

Substituting $\langle k \rangle = \frac{2rc}{1+r}$ into Eq. (17), then get:

$$\frac{\partial g}{\partial z} - \frac{\frac{2r}{1+r}}{(1-z)(\frac{2}{1+r}-z)}g(z) = \frac{\frac{2r}{1+r}z^c}{(1-z)(z-\frac{2}{1+r})} \quad (18)$$

If $r=1$, which means that the network of debtor-creditor relationships is a constant-size network. We didn't discuss this special because it was discussed by Moore et al.(2006) and it was inconsistent with the actual P2P lending market.

If $r \neq 1$, the Eq. (18) is an ordinary differential equation, and the general solution of Eq. (18) is

$$g(z) = Be^{\int\frac{\frac{2r}{1+r}}{(1-z)(\frac{2}{1+r}-z)}dz} + e^{\int\frac{\frac{2r}{1+r}}{(1-z)(\frac{2}{1+r}-z)}dz}\int\frac{rz^c}{(1-z)(\frac{z(1+r)}{2}-1)}e^{-\int\frac{\frac{2r}{1+r}}{(1-z)(\frac{2}{1+r}-z)}dz}dz \quad (19)$$

One of integrating factors is:

$$\int\frac{\frac{2r}{1+r}}{(1-z)(\frac{2}{1+r}-z)}dz = \int\left[\frac{\frac{2r}{r-1}}{1-Z} - \frac{\frac{2r}{r-1}}{\frac{2}{r+1}-z}\right]dz = \frac{2r}{r-1}\ln\frac{|\frac{2}{r+1}-z|}{|1-z|} \quad (20)$$

If $r > 1, \alpha = \frac{2}{r+1} < 1$, then get a special solution in $[\alpha, z]$:

$$g(z) = \frac{2r}{1+r}\left(\frac{z-\alpha}{1-z}\right)^{-\frac{2r}{r-1}} \times \int_\alpha^z \frac{t^c}{(1-t)(t-\alpha)}\left(\frac{\alpha-t}{1-t}\right)^{\frac{2r}{r-1}}dt \tag{21}$$

Setting:

$$u = \frac{\alpha - t}{1 - \alpha} \quad t = \alpha - (1-\alpha)u$$

And Substituting $u$ and $t$ into Eq. (21), then get:

$$g(z) = \frac{2r}{1+r}\left(\frac{z-\alpha}{1-z}\right)^{\frac{2r}{r-1}} \times \int_0^{\frac{\alpha-z}{1-\alpha}} \frac{(\alpha-(1-\alpha)u)^c}{(1-\alpha+(1-\alpha)u)(\alpha-(1-\alpha)u-\alpha)}\left(\frac{\alpha-\alpha+(1-\alpha)u}{1-\alpha+(1-\alpha)u}\right)^{\frac{2r}{1-r}}(-(1-\alpha))du$$

$$g(z) = \frac{2r}{1+r}\left(\frac{z-\alpha}{1-z}\right)^{-\frac{2r}{r-1}} \times \int_0^{\frac{\alpha-z}{1-\alpha}} \frac{(\alpha-(1-\alpha)u)^c}{(1-\alpha)(1+u)u}\left(\frac{u}{1+u}\right)^{\frac{2r}{r-1}}du$$

$$g(z) = \frac{2r}{1+r}\left(\frac{a-z}{1-z}\right)^{-\frac{2r}{r-1}} \times \int_0^{\frac{\alpha-z}{1-\alpha}} \frac{(\alpha-(1-\alpha)u)^c}{(1-\alpha)u^2}\frac{u}{1+u}\left(\frac{u}{1+u}\right)^{\frac{2r}{r-1}}du$$

$$g(z) = \frac{2r}{1+r}\left(\frac{a-z}{1-z}\right)^{-\frac{2r}{r-1}} \times \int_0^{\frac{\alpha-z}{1-\alpha}} \frac{(\alpha-(1-\alpha)u)^c}{(1-\alpha)u^2}\left(\frac{u}{1+u}\right)^{\frac{3r-1}{r-1}}du$$

Setting:

$$\gamma = \frac{3r-1}{r-1}$$

Then get:

$$g(z) = \frac{2r}{1+r}\left(\frac{a-z}{1-z}\right)^{1-\gamma} \times \int_0^{\frac{\alpha-z}{1-\alpha}} \frac{(\alpha-(1-\alpha)u)^c}{(1-\alpha)u^2}\left(\frac{u}{1+u}\right)^\gamma du \tag{22}$$

The exact solutions of such as the Eq.(22) was discussed by Moore et al. (2006), and the degree distribution was a power-law with exponent γ:

$$p_k \sim (\gamma-1)k^{-\gamma}\int_0^\infty e^{-(1-\alpha)x}x^{\theta-2}dx = \frac{\Gamma(\gamma)}{(1-\alpha)^{\gamma-1}}k^{-\gamma} \tag{23}$$

In Eq.(23), $\Gamma(\gamma) = \int_0^\infty e^{-x}x^{\gamma-1}dx$ is $\Gamma$-function. In the limit $r \to \infty$, the network of debtor-creditor relationships is power-law behavior $p_k \sim k^{-\gamma}$ with preferential attachment.

## 3 A Case of P2P lending

### 3.1 Description of the Data

Renrendai.com has rapidly grown into one of the largest P2P lending market in China. By February 2018, Renrendai.com had facilitated over $7.8 billion in loans[2]. Our dataset were the samples of listing posted on Renrendai from website's inception in October 2010 through December 2014, but it don't include those loan originations which had not been funded. Fig.4 shows the network of debtor-creditor relationship by 31 October 2010.

The yellow nodes represents lender, blue nodes represents borrowers and red nodes represents someone who was a borrower of loan originations was a lender of other loans. The number in the node was unique identification of participant on Renrendai.com. Because we focused on the existing of debtor-creditor relationships, for convenient study, we ignored the question that who was borrower or lender.

---

[2] www.renrendai.com

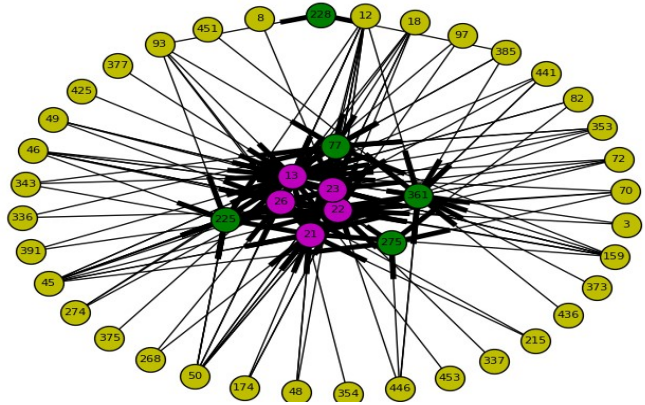

Fig.4: The network of debtor-creditor relationship by 31 October

By 31 December 2014, there were 151759 nodes and 3716699 edges in the network of debtor-creditor relationships. The data in Fig.5 shows the number of nodes in the network of debtor-creditor relationships, and the data in Fig.6 shows the number of edges in the network of debtor-creditor relationships. The data in Fig.7 shows the number of added nodes, and the data in Fig.8 shows the number of added edges. In Fig.5~Fig.8, there was a clear trend of increasing in debtor-creditor relationships network since 2013, this trend of increasing due to P2P lending market rapidly expanded in China since 2013.

Suppose that $\frac{n_k}{n}$ was the actual fraction of nodes that have degree $k$ in the network of debtor-creditor relationships at a given time. According to $p_k \sim k^{-\gamma}$, setting $k^{-\gamma} = \frac{n_k}{n}$, then

$$ln\frac{n_k}{n} = -\gamma \ln k \tag{24}$$

Based on above equation and the actual data of degree distribution at every day, the exponent of power-law degree distributions $\gamma$ can be calculated by OLS method (in Fig.11). The data in Fig.11 shows that there was a clear trend of increasing while the number of nodes represents the borrowers or lenders rapidly increased in 2013. But the exponent of power-law $\gamma$ didn't increased rapidly while the number of nodes rapidly increased yet in 2014. Then the data in Fig.9 shows the exponent of power-law degree distributions $\gamma$ may be approximately in the range of 1.84-1.93. The data in Table 1 shows the simple summary statistics about the exponent of power-law $\gamma$.

The data in Fig.9 and Fig.10 shows the number of deleted nodes and edges due to the expired loans. The data in Fig.11 shows the number of degree of whole network and the average degree of added nodes in each day. It was found that the average degree of whole network may be approximately 40. From Table 3,we can see that the average degree of whole network in 2013 was less than average degree of whole network in 2012, because the increasing rate of nodes was less than the increasing rate of edges in 2013 while the increasing rate of nodes more than the increasing rate of edges in 2012(Table 2). More added nodes does not correspond with more added edges means that although there were more and more persons take parted in P2P lending market, the lending-borrowing activities of each participant may be downed.

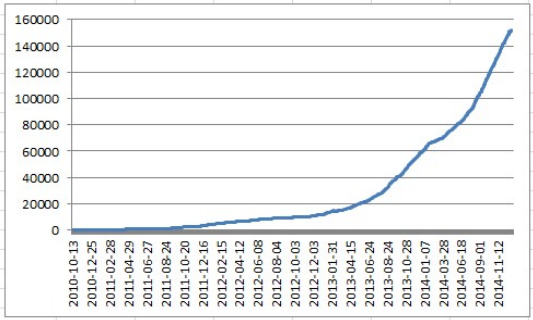
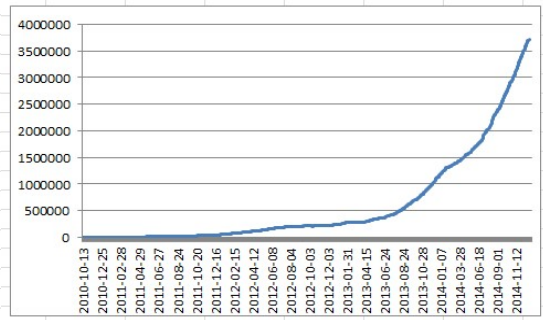

Fig.5: The number of vertices in debtor-creditor relationships network

Fig.6: The number of edges in debtor-creditor relationships

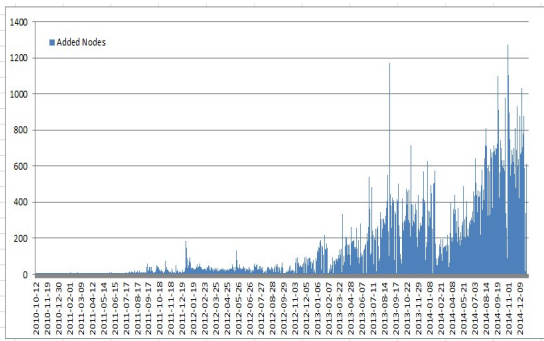
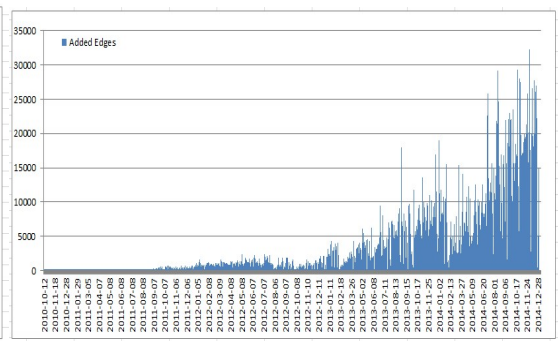

Fig.7: The number of added nodes in debtor-creditor relationships network

Fig.8: The number of added edged in debtor-creditor relationships network

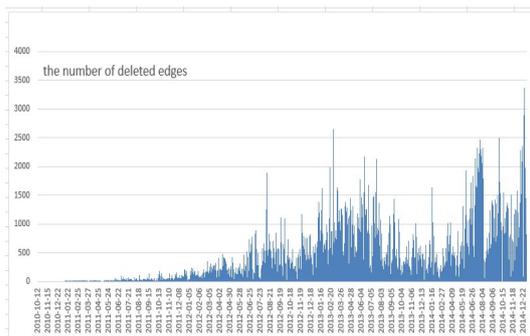
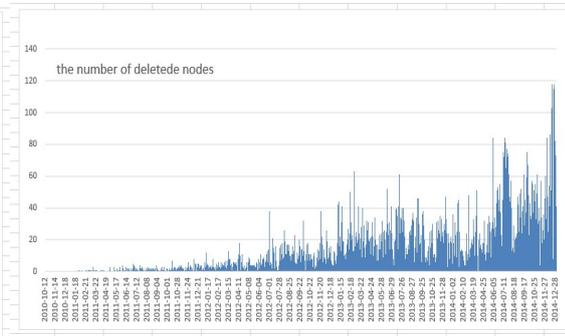

Fig.9: The number of deleted nodes in debtor-creditor relationships network

Fig.10: The number of deleted edges in debtor-creditor relationships

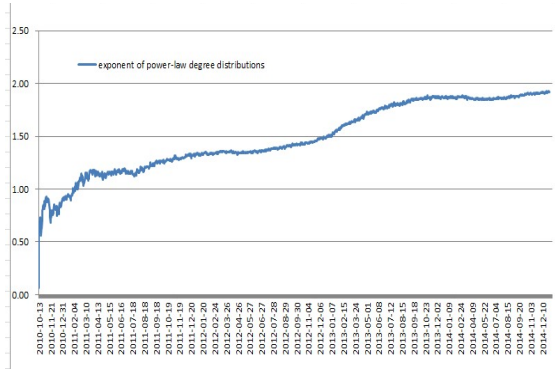
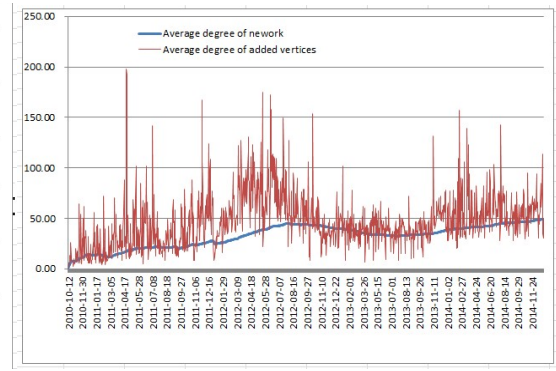

Fig.11: The exponent of power-law degree distributions

Fig.12: The average degree of debtor-creditor relationships network

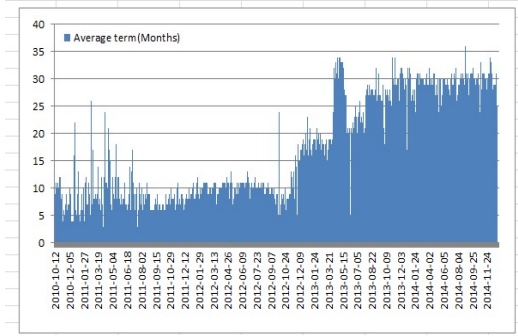
Fig.13: The average term of loan

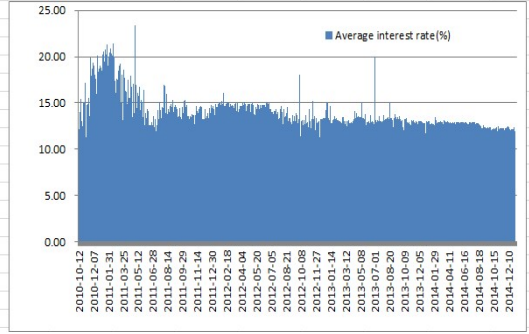
Fig.14: The average interest rate of loan

Table 1 The exponent of power-law

| Year | Min | Max | mean | St. dev |
|---|---|---|---|---|
| 2010 | 0.0648 | 0.9296 | 0.7900 | 0.1249 |
| 2011 | 0.8935 | 1.3369 | 1.1730 | 0.1057 |
| 2012 | 1.3151 | 1.5144 | 1.3842 | 0.0478 |
| 2013 | 1.4994 | 1.8848 | 1.7503 | 0.1091 |
| 2014 | 1.8427 | 1.9269 | 1.8759 | 0.0219 |

Table 2 Number of nodes and edges

| Date | Number of nodes | Number of edges | The rate of increasing nodes | The rate of increasing edges |
|---|---|---|---|---|
| 2010/12/31 | 151 | 1048 | --- | ---- |
| 2011/12/31 | 3483 | 45348 | 2306.62% | 4327.10% |
| 2012/12/31 | 11850 | 236115 | 340.22% | 520.67% |
| 2013/12/31 | 59330 | 1137474 | 500.68% | 481.75% |
| 2014/12/31 | 151759 | 3716699 | 255.79% | 326.75% |

Table 3 Average degree of network

| Year | Min | Max | mean | Std. Dev. |
|---|---|---|---|---|
| 2010 | 1.00 | 14.26 | 9.80 | 3.2176 |
| 2011 | 11.68 | 27.78 | 19.59 | 4.2828 |
| 2012 | 24.76 | 44.65 | 37.94 | 6.3267 |
| 2013 | 32.33 | 39.83 | 35.07 | 2.1079 |
| 2014 | 38.34 | 49.04 | 43.67 | 3.0985 |

### 3.2 Factors of influencing the exponent of power-law

Although there are a lot of literatures studying network with the power-law distributions of degree, little attention has been paid to the factors of influencing the exponent of power-law.

Refer to Eq. (24), it is known that $\gamma$ is relevant to $\ln(n)$. Beyond this, we consider that the properties of loans have some impact on $\gamma$. For example, lenders preferentially to make a loan with higher interest rate or lower risk, as a result, loan originations with higher return or lower risk will attract more lenders to lend them money. In other words, loan originations with higher return or lower risk may lead to more edges represented the creditor-debtor relationship. In order to test whether the properties of loans affect exponent of power-law, the data in Fig.11 denoted by $\gamma_t$ was the dependent variable and the average interest rate denoted by $R_t$ and average term denoted by $M_t$ were independent variables respectively. The data in Fig.13 was $M_t$, and the data in Fig.14 was $R_t$. $\ln(n_t)$ was controlling variable, and $n_t$ is the number of nodes each day. Finally the model was:

$$\gamma_t = c + \alpha R_t + \beta M_t + \varphi \ln(n_t) + \varepsilon_t \tag{25}$$

Finally the number of our sample were 1353 and date were in 12 October 2010 through 31 December 2014. Because there weren't new loan originations on web of P2P lending platform at some days, this lead to the $R_t$ and $M_t$ were equal to 0 on those days, so we don't choose those data that $R_t$ and $M_t$ were equal to 0. Table 4 presented the summary statistics for our sample and Table 5 presented the Pearson correlations for our sample.

Table 4 Summary Statistics of variables

| variable | Mean | Median | Maximum | Minimum | Std. Dev. |
|---|---|---|---|---|---|
| $\gamma_t$ | 1.49 | 1.42 | 1.93 | 0.06 | 0.32 |
| $n_t$ | 28367.87 | 9557 | 151759 | 5 | 38207.59 |
| $R_t$ | 0.1355 | 0.1306 | 0.234 | 0.09 | 0.018 |
| $M_t$ | 15.35 | 11 | 36 | 3 | 9.6 |

Number of observations: 1353

Table 5 Pearson correlations of variables

| Correlation | $\gamma_t$ | $R_t$ | $n_t$ | $M_t$ |
|---|---|---|---|---|
| $\gamma_t$ | 1.000000 | | | |
| $R_t$ | -0.535755 | 1.000000 | | |
| $n_t$ | 0.767023 | -0.386482 | 1.000000 | |
| $M_t$ | 0.832488 | -0.319328 | 0.781582 | 1.000000 |

Finally to estimate Eq. (25) by OLS method and got:

$$\gamma_t^* = 0.368 - 0.701 R_t + 0.0074 M_t + 0.124 \ln(n_t) \tag{26}$$
$$(-5.962)^{***} \quad (25.894)^{***} \quad (80.566)^{***}$$

In Eq.(25), the coefficient of $R_t$ shows that the exponent of power-law is negatively correlated with the interest rate. Given a degree $k$, the coefficient of $R_t$ shows that higher interest rate leads to lower exponent of power-law and $k^{-\gamma}$ is larger, $k^{-\gamma}$ represents the probability that a debtor has $k$ creditor-debtor relationships. In other words, lenders preferentially choose the borrowers who give higher interest rate of loan originations. The coefficient of $M_t$ shows that the exponent of power-law is positively correlated with the term of loans, which means that longer-term lead to higher exponent of power-law and $k^{-\gamma}$ is smaller. In other words, lenders preferentially choose the borrower who give shorter-term of loan originations. In short, these results are coincident with the risk characteristic of P2P lending market such as higher default risk lead to lender to demand higher return and higher uncertainty lead to lender to choose short-term of investing.

## 4 Conclusion

In this paper we have studied the networks of debtor-creditor relationships in P2P lending market. According to the loan attributes such as there must exist a date on which debtors must pay off the principal and interest, and after debtors do it, the debtor-creditor relationships are cleaned. In addition, consider new debtor-creditor relationships have come into being continually due to new loan originations, the networks of debtor-creditor relationships evolve with addition and deletion of nodes. In short, these actual lending activities provide a case for theory model of evolving network.

Consider the loan attributes and based on a different assumption from Moore et al. (2006), yet getting the same conclusion that the degree distribution of an evolving network with addition and deletion of nodes is satisfied with power-law distribution under preferential attachment. In our case, the result of empirical study is consistent in power-law distribution, and the exponent of power-law of degree distributions γ may be approximately in the range of 1.84~1.93. On the other hand, as Moore et al. (2006) have commented, although the exponent of power-law generated by theoretical model appear not to be in agreement with the result in our case. There are good reasons to believe that the networks of debtor-creditor relationships are scale-free networks if it appears a net growth network.

In addition, we study what factors impact on the exponent of power-law besides the number of nodes. We find that the both interest rate and term have significantly influence on the exponent of power-law degree distributions. Interest rate is negatively correlated with the exponent of power-law and term is positively correlated with the exponent of power-law. This result is significant because most previous studying of network don't care for what factors impact on the exponent of power-law.

## Acknowledgments

This work was funded by The National Natural Science Foundation of China (NSFC) (Project ID: 71571030).